\RequirePackage{ifpdf}
\ifpdf 
\documentclass[pdftex]{sigma}
\else
\documentclass{sigma}
\fi

\begin{document}

\allowdisplaybreaks

\renewcommand{\thefootnote}{$\star$}

\renewcommand{\PaperNumber}{089}

\FirstPageHeading

\ShortArticleName{Junction Type Representations of the Temperley--Lieb Algebra}

\ArticleName{Junction Type Representations of the Temperley--Lieb Algebra and Associated Symmetries\footnote{This paper is a
contribution to the Proceedings of the International Workshop ``Recent Advances in Quantum Integrable Systems''. The
full collection is available at
\href{http://www.emis.de/journals/SIGMA/RAQIS2010.html}{http://www.emis.de/journals/SIGMA/RAQIS2010.html}}}

\Author{Anastasia DOIKOU~$^\dag$ and Nikos KARAISKOS~$^{\dag\ddag}$}

\AuthorNameForHeading{A.~Doikou and N. Karaiskos}

\Address{$^\dag$~Department of Engineering Sciences, University of Patras, GR-26500 Patras, Greece}
\EmailD{\href{mailto:adoikou@upatras.gr}{adoikou@upatras.gr}, \href{mailto:nkaraiskos@upatras.gr}{nkaraiskos@upatras.gr}}

\Address{$^\ddag$~Centre de Physique Th\'eorique, Ecole Polytechnique, CNRS-UMR 7644,\\
\hphantom{$^\ddag$}~91128 Palaiseau, France}

\ArticleDates{Received September 06, 2010, in f\/inal form November 29, 2010;  Published online December 07, 2010}

\Abstract{Inspired by earlier works on representations of the Temperley--Lieb algebra we introduce a novel family of representations
of the algebra. This may be seen as a generalization of the so called asymmetric twin representation.
The underlying symmetry algebra is also examined and it is shown that in addition to certain obvious exact quantum
symmetries non trivial quantum algebraic realizations that exactly commute with the representation also exist.
Non trivial representations of the boundary Temperley--Lieb algebra as well as the related residual symmetries
are also discussed. The corresponding novel $R$ and $K$ matrices solutions of the Yang--Baxter and ref\/lection equations
are identif\/ied, the relevant quantum spin chain is also constructed and its exact symmetries are studied.}


\Classification{81R50; 17B37; 17B80}

\renewcommand{\thefootnote}{\arabic{footnote}}
\setcounter{footnote}{0}

\section{Introduction}

There has been an increasing activity in recent years on the study of representations of
(af\/f\/ine) Hecke algebras \cite{hecke,ahecke, cherednik, dema, doikoutwin, doikoumartin, doikouhecke, fradkin, graham, isog, jimbo, jimbo0, hecke1, kuma, levymartin, martinwood, martinpotts, mene, nichols1, hecke5} and their quotients, which can provide solutions \cite{jimbo,jimbo0, hecke1, kuma, levymartin, martinwood, martinpotts, mene,nichols1} of the Yang--Baxter \cite{baxter} and the ref\/lection equations \cite{cherednik, sklyanin}. Such investigations are of great physical signif\/icance due to the fact that the obtained solutions can be implemented into physical integrable systems, and conformal f\/ield theories with open boundary conditions that preserve integrability (see e.g.~\cite{doikoumartin, doikouhecke, kuma, nichols1,  cft}), or lead to the construction of novel physical systems as will be transparent in the present investigation (see also \cite{doikoumartin}).
In the present article, we focus on the investigation of novel representations in the context of the
Temperley--Lieb (TL) algebra~\cite{tl, martinpotts},
the so called junction representation. In fact, this may be seen as a natural generalization of the asymmetric twin
(or cable) representation studied in~\cite{martinwood, doikoumartin, doikoutwin}. More precisely, the representation is introduced and it is shown that it satisf\/ies the def\/ining
relations of the algebra, provided that certain constraints are satisf\/ied. In addition, the existence of non trivial quantum
algebras that are exact symmetries of the aforementioned representation is investigated. Our analysis is further extended in the case of the boundary
Temperley--Lieb (blob) algebra,
and non trivial representations for the ``boundary'' element of the algebra are identif\/ied.

It is worth noting that the representation leads naturally to novel physical integrable systems~-- quantum spin chains,
with non-trivial integrable boundary conditions. It is thus clear that the whole analysis presents not only mathematical, but
physical interest as well, given that such investigations provide new boundary conditions that
may alter the physical behavior as well as the symmetries associated to the corresponding system (see e.g.~\cite{doikoumartin, doikouhecke, doikoutwin}).

The outline of this article is as follows: in the next section the def\/initions of the Temperley--Lieb algebra are reviewed,
the junction representation is introduced, and it is shown that it satisf\/ies the Temperley--Lieb algebra def\/ining relations.
The exact symmetries
associated to the junction
representation are then examined, certain obvious and non-trivial representations of quantum algebras are introduced, and  turn out to exactly commute with
all the elements of the Temperley--Lieb algebra in the junction representation.
Explicit proof for the f\/irst non trivial case is provided, and expressions of a whole family of representations of
the quantum algebra for the generic case are conjectured. In Section~\ref{section3} the boundary Temperley--Lieb algebras
and the related representations are discussed, and the residual symmetries are examined. In the last section we introduce the relevant quantum spin chain, and we also examine the
asso\-cia\-ted symmetries of the open transfer matrix.
In the Appendices various technical points are illustrated.

\section[Temperley-Lieb algebra and representations]{Temperley--Lieb algebra and representations}\label{section2}

Let us brief\/ly review the basic notions associated to the Temperley--Lieb algebra \cite{tl, martinpotts}.
The Temperley--Lieb $T_N(q)$ algebra is def\/ined by the generators ${\mathbb U}_i$,
$i =1, \dots, N$ satisfying
\begin{gather}
{\mathbb U}_i^2 = -\big(q+q^{-1}\big){\mathbb U}_i, \qquad
{\mathbb U}_{i} {\mathbb U}_{i\pm 1}  {\mathbb U}_{i} = {\mathbb U}_i, \nonumber\\
[{\mathbb U}_i,  {\mathbb U}_j] =0, \qquad |i-j|>1.\label{tl}
\end{gather}
A well known representation of the Temperley--Lieb algebra is the XXZ one.
More precisely, consider the matrix
\begin{gather}
U = \sum_{a \neq b =1}^2 e_{ab} \otimes e_{ba} - \sum_{a \neq b =1}^2 q^{-{\rm sgn}(a-b)} e_{aa} \otimes e_{bb},
\label{U_XXZ}
\end{gather}
where the matrices $e_{ab}$ are in this case $2\times 2$  matrices def\/ined as: $(e_{ab})_{cd} = \delta_{ac}  \delta_{bd}$.
Then take the representation of the Temperley--Lieb algebra
$\rho:  T_n(q) \mapsto \mbox{End}(({\mathbb C}^2)^{\otimes N})$ such that
\[
\rho({\mathbb U}_i) = {\mathbb I} \otimes \cdots \otimes {\mathbb I} \otimes \underbrace{U}_{i,\; i+1}
\otimes {\mathbb I}\otimes
\cdots \otimes {\mathbb I}.
\]

Inspired by the asymmetric twin (or cable) representation \cite{martinwood, doikoumartin}
of the Temperley--Lieb algebra we
introduce a novel representation, which we shall call for obvious reasons the
{\it junction representation}.
This involves essentially $n$ copies of the XXZ representation i.e.\
$\Theta: T_n(q) \mapsto \mbox{End}((({\mathbb C}^2)^{\otimes n})^{\otimes N})$
\begin{gather}
\Theta({\mathbb U}_l) =\prod_{i=1}^n \rho_{a_i}({\mathbb U}_{l^{(i)}}), \label{product}
\end{gather}
where $a_i$ is the parameter associated with the $i$-th copy of the Temperley--Lieb algebra
and is a~representation of the Temperley--Lieb algebra itself provided that:
\begin{gather}
(-1)^n\prod_{i=1}^n \big(a_i +a_i^{-1}\big) = - \big(q+q^{-1}\big).
\label{cond_for_a}
\end{gather}
The latter condition apparently follows from the quadratic relation (Hecke condition) of the Temperley--Lieb algebra.
Now observe that the matrices $\rho_{a_i}({\mathbb U}_{l^{(i)}})$ commute among each other,
since they act on dif\/ferent vector spaces. Having that in mind and taking into account that
each one of the matrices satisfy the TL algebra,
one can immediately show that $\Theta$ provides a~representation of the TL algebra. Constraints
are only entailed from the condition~\eqref{cond_for_a}. One then obtains $n$ equations with $n$ unknown quantities (see Appendix~\ref{appendixA} for more details on the Hecke condition), hence $a_i$'s can be explicitly determined. Recall that the asymmetric twin representation involves two copies of the XXZ representation i.e.~$n=2$.
Note that we consider here the following sequence of spaces in the tensor product:
$\cdots V_{l^{(1)}} \otimes V_{l^{(2)}} \otimes \cdots \otimes V_{l^{(n)}} \otimes
V_{(l+1)^{(1)}}\otimes V_{(l+1)^{(2)}}\otimes \cdots \otimes V_{(l+1)^{(n)}} \cdots$, $V \equiv {\mathbb C}^2.$
It is clear that the total number of ${\mathbb C}^2$ spaces (i.e.\ total number of ``sites'') is $nN$.

The junction representation can be schematically represented as follows:
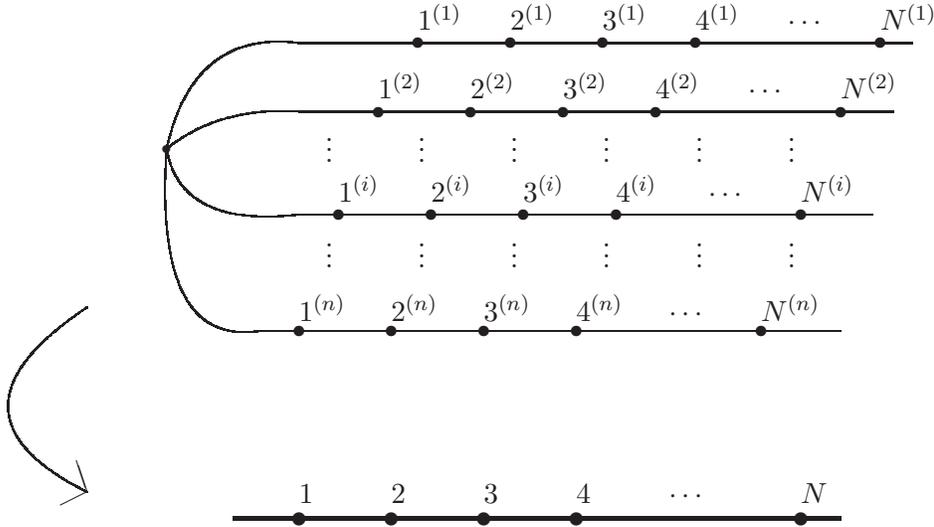
\begin{center}
\begin{picture}(1,120)(180,-60)
\qbezier(30,0)(40,45)(80,40)
\qbezier(30,0)(50,16)(80,14)
\qbezier(30,0)(35,-30)(80,-25)
\qbezier(30,0)(25,-75)(65,-69)
\put(90,-5){$\vdots$} \put(125,-5){$\vdots$} \put(160,-5){$\vdots$}
\put(195,-5){$\vdots$} \put(230,-5){$\vdots$} \put(265,-5){$\vdots$}
\put(90,-45){$\vdots$} \put(125,-45){$\vdots$} \put(160,-45){$\vdots$}
\put(195,-45){$\vdots$} \put(230,-45){$\vdots$} \put(265,-45){$\vdots$}
\put(80,40){\line(1,0){232}} \put(80,14){\line(1,0){225}}
\put(80,-25){\line(1,0){217}} \put(65,-69){\line(1,0){220}}
\put(125,40){\circle*{4}} \put(160,40){\circle*{4}}
\put(195,40){\circle*{4}} \put(230,40){\circle*{4}} \put(300,40){\circle*{4}}
\put(110,14){\circle*{4}} \put(145,14){\circle*{4}}
\put(180,14){\circle*{4}} \put(215,14){\circle*{4}} \put(285,14){\circle*{4}}
\put(95,-25){\circle*{4}} \put(130,-25){\circle*{4}}
\put(165,-25){\circle*{4}} \put(200,-25){\circle*{4}} \put(270,-25){\circle*{4}}
\put(80,-69){\circle*{4}} \put(115,-69){\circle*{4}}
\put(150,-69){\circle*{4}} \put(185,-69){\circle*{4}} \put(255,-69){\circle*{4}}
\put(125,45){$1^{(1)}$} \put(160,45){$2^{(1)}$}
\put(195,45){$3^{(1)}$} \put(230,45){$4^{(1)}$}
\put(265,45){$\cdots$} \put(300,45){$N^{(1)}$}
\put(110,19){$1^{(2)}$} \put(145,19){$2^{(2)}$}
\put(180,19){$3^{(2)}$} \put(215,19){$4^{(2)}$}
\put(250,19){$\cdots$} \put(285,19){$N^{(2)}$}
\put(95,-20){$1^{(i)}$} \put(130,-20){$2^{(i)}$}
\put(165,-20){$3^{(i)}$} \put(200,-20){$4^{(i)}$}
\put(235,-20){$\cdots$} \put(270,-20){$N^{(i)}$}
\put(80,-65){$1^{(n)}$} \put(115,-65){$2^{(n)}$}
\put(150,-65){$3^{(n)}$} \put(185,-65){$4^{(n)}$}
\put(220,-65){$\cdots$} \put(255,-65){$N^{(n)}$}
\put(30,0){\circle*{3}}

\qbezier(0,-60)(-60,-100)(0,-130)
\put(0,-130){\line(-2,-1){10}}
\put(0,-130){\line(-1,3){4}}

\linethickness{0.5mm}
\put(55,-140){\line(1,0){230}}
\put(80,-140){\circle*{5}} \put(115,-140){\circle*{5}}
\put(150,-140){\circle*{5}} \put(185,-140){\circle*{5}}
\put(270,-140){\circle*{5}}
\put(80,-134){$1$} \put(115,-134){$2$}
\put(150,-134){$3$} \put(185,-134){$4$}
\put(220,-134){$\cdots$} \put(270,-134){$N$}

\end{picture}\vspace{35mm}
\end{center}

The thick line schematically presents the
junction representation, showing somehow the ``fusion'' of $n$ copies of the XXZ
representation. In particular, each one of the indices $l$ appea\-ring there may be seen as a
``fusion'' of the indices $l^{(i)}$ as: $~l \equiv \langle l^{(1)}\,  l^{(2)} \ldots\, l^{(N)} \rangle$, $l+1 \equiv \langle (l+1)^{(1)} \, (l+1)^{(2)} \ldots \, (l+1)^{(N)} \rangle$ and so on.

Let us now explicitly express each one of the terms in the product \eqref{product}, def\/ining the junction representation:
\begin{gather*}
\rho_{a_m}({\mathbb U}_{l^{(m)}} ) = \cdots{\mathbb I} \otimes \left(\sum_{a \neq b}
 \underbrace{{\mathbb I} \otimes \cdots \otimes {\mathbb I}}_{m-1} \otimes \underbrace{e_{ab}}_{l^{(m)}}
 \otimes \underbrace
{{\mathbb I}\otimes\cdots \otimes{\mathbb I} }_{n-1} \otimes  \underbrace{e_{ba}}_{(l+1)^{(m)}} \otimes
\underbrace{{\mathbb I}\otimes \cdots \otimes {\mathbb I}}_{n-m}\right.  \nonumber\\
\qquad {} - \left.\sum_{a\neq b}
a_m^{-{\rm sgn}(a-b)}  \underbrace{{\mathbb I} \otimes \cdots \otimes {\mathbb I}}_{m-1} \otimes  \underbrace{e_{aa}}_{l^{(m)}}
 \otimes \underbrace
{{\mathbb I}\otimes\cdots \otimes{\mathbb I} }_{n-1} \otimes \underbrace{e_{bb}}_{(l+1)^{(m)}} \otimes
\underbrace{{\mathbb I}\otimes \cdots \otimes {\mathbb I}}_{n-m}
\right)\otimes {\mathbb I}\cdots \nonumber\\ \textrm{with} ~ m \in \{1, \ldots, n \}.
\end{gather*}
The matrix above clearly acts non-trivially on the spaces
${\mathbb C}^2_{l^{(m)}}\otimes {\mathbb C}^2_{(l+1)^{(m)}}$.

\subsection[The symmetry: $n=3$]{The symmetry: $\boldsymbol{n=3}$}\label{section2.1}

For the sake of simplicity and for illustrating purposes, we shall focus in this section on the simplest non-trivial case where $n=3$. Nevertheless, expressions on the symmetries associated to higher $n$ will be presented in the next subsection.

The junction representation of the Temperley--Lieb algebra is in this case
$\Theta:  T_N(q) \mapsto \mbox{End}(({\mathbb C}^2\otimes {\mathbb C}^2\otimes {\mathbb C}^2)^{\otimes N})$
\[
\Theta({\mathbb U}_l) = \rho_{a_1}({\mathbb U}_{l^{(1)}})\,  \rho_{a_2}({\mathbb U}_{l^{(2)}})\,
\rho_{a_3}({\mathbb U}_{l^{(3)}}).
\]
It is clear from their structure that the above matrices $\rho_{a_i}({\mathbb U}_{l^{(i)}})$ act on
dif\/ferent spaces and hence they commute with each other.
Let us now come to the study of the exact quantum symmetry algebras associated to the junction representation.

Let us f\/irst brief\/ly recall the $U_q(sl_2)$ algebra, def\/ined by generators $e$, $f$, $h$ that satisfy \cite{jimbo0}:
\[
[h, e] = 2e,  \qquad [h, f] = -2f ,\qquad [e, f] = \frac{q^h -q^{-h}}{q -q ^{-1}}.
\]
The quantum algebra is as known equipped with a non-trivial co-product
$\Delta: ~U_q(sl_2) \mapsto U_q(sl_2) \otimes U_q(sl_2)$
\[
\Delta(q^h)= q^h\otimes q^h, \qquad \Delta(x) = x \otimes q^{\frac{h}{2}} + q^{-\frac{h}{2}} \otimes x,
\qquad x \in \{e,  f\}.
\]
The $n$ co-product may be obtained by iteration as: $\Delta^{(n)} = ({\mathbb I} \otimes \Delta^{(n-1)})\Delta$.

Recall that the spin $\frac{1}{2}$ representation of the algebra above is given by, $\pi_q: U_q(sl_2)
\mapsto \mbox{End}({\mathbb C}^2)$
\[
\pi_q(h) = \sigma^z, \qquad \pi_q(e) = \sigma^+, \qquad \pi_q(f) = \sigma^-,
\]
where $\sigma^z$, $\sigma^{\pm}$ are the $2\times 2$ Pauli matrices. Now def\/ine:
\begin{gather*}
\pi_1(x) = \pi_{a_1}(x) \otimes {\mathbb I} \otimes {\mathbb I},
\qquad \pi_2(x) =  {\mathbb I}\otimes \pi_{a_2}(x) \otimes {\mathbb I},
\\
\pi_3(x) =  {\mathbb I}\otimes  {\mathbb I} \otimes \pi_{a_3}(x), \qquad x \in U_{a_i}(sl_2).
\end{gather*}
Concerning the junction representation, it is easy to check that:
\begin{gather}
\big [\Theta({\mathbb U}_l),  \pi_i^{\otimes N}(\Delta^{(N)}(x)) \big ] =0 , \qquad x \in U_{a_i}(sl_2),\qquad i=1,
  2,  3. \label{basic}
\end{gather}
It is obvious that each one of the representations acts on the spaces $V_{l^{(i)}}$ separately, that is
\[
\big[\rho_{a_i}(U_{l^{(i)}}),  \pi_j^{\otimes N}(\Delta^{(N)}(x))\big] =0, \qquad i,  j \in \{1,  2,  3\},
\]
which immediately leads to \eqref{basic}. Hence, the junction representation enjoys a manifest
\begin{gather*}
{\cal G}_0 \equiv U_{a_1}(sl_2)\otimes U_{a_2}(sl_2)\otimes U_{a_3}(sl_2)
\end{gather*}
symmetry, as is trivially expected by its construction.

However, as in the case of the twin asymmetric representation \cite{doikoumartin, doikoutwin}, our main objective is to
extract non trivial quantum algebra realizations that commute with the junction representation.
Consider the following representations ${\mathrm f}_{i} : U_{q_i}(sl_2) \mapsto
\mbox{End}(({\mathbb C}^{(2)})^{\otimes 3})$, $i=0,1,2,3$
\begin{subequations}\label{rep}
\begin{gather}
{\mathrm f}_{0}(h) =  e_{11} \otimes e_{11} \otimes e_{11}  -e_{22}\otimes
e_{22} \otimes e_{22} = h_1^{(0)} - h_2^{(0)},
\nonumber\\
{\mathrm f}_{0}(e) = e_{12} \otimes e_{12} \otimes e_{12}, \qquad
{\mathrm f}_{0}(f) = e_{21} \otimes e_{21} \otimes e_{21}, \nonumber\\
{\mathrm f}_{0}(q_0^h) = {\mathbb I} + (q_0-1)h_1^{(0)} + \big(q_0^{-1} -1\big)h_2^{(0)},\label{rep0}
\\
{\mathrm f}_{1}(h) =  e_{22}\otimes
e_{11} \otimes e_{11} -e_{11} \otimes e_{22} \otimes e_{22} = h_1^{(1)} - h_2^{(1)},
\nonumber\\
{\mathrm f}_{1}(e) = e_{21} \otimes e_{12} \otimes e_{12}, \qquad
{\mathrm f}_{1}(f) = e_{12} \otimes e_{21} \otimes e_{21}, \nonumber\\
{\mathrm f}_{1}(q_1^h) = {\mathbb I} + (q_1-1)h_1^{(1)} + \big(q_1^{-1} -1\big)h_2^{(1)},\label{rep01}
\\
{\mathrm f}_{2}(h) = e_{11} \otimes e_{22} \otimes e_{11} -e_{22}\otimes
e_{11} \otimes e_{22} = h_1^{(2)} - h_2^{(2)},
\nonumber\\
{\mathrm f}_{2}(e) = e_{12} \otimes e_{21} \otimes e_{12}, \qquad
{\mathrm f}_{2}(f) = e_{21} \otimes e_{12} \otimes e_{21}, \nonumber\\
{\mathrm f}_{2}(q_2^h) = {\mathbb I} + (q_2-1)h_1^{(2)} + \big(q_2^{-1} -1\big)h_2^{(2)},\label{rep02}
\\
{\mathrm f}_{3}(h) =  e_{11} \otimes e_{11} \otimes e_{22}-e_{22}\otimes
e_{22} \otimes e_{11} = h_1^{(3)} - h_2^{(3)},
\nonumber\\
{\mathrm f}_{3}(e) = e_{12} \otimes e_{12} \otimes e_{21}, \qquad
{\mathrm f}_{3}(f) = e_{21} \otimes e_{21} \otimes e_{12},\nonumber\\
{\mathrm f}_{3}(q_3^h) = {\mathbb I} + (q_3-1)h_1^{(3)} + \big(q_3^{-1} -1\big)h_2^{(3)}. \label{rep03}
\end{gather}
\end{subequations}
It is straightforward to check that each one of the $\mathrm{f}_i$'s def\/ined above provide a representation of~$U_{q_i}(sl_2)$.

After some quite cumbersome but straightforward computation, it is shown that
\begin{gather}
\big[\Theta({\mathbb U}_l),  {\mathrm f}_{i}^{\otimes N}(\Delta^{(N)}(x))\big] =0, \qquad x \in U_{q_i}(sl_2),
\qquad i=0,1,2,3, \label{symm0}
\end{gather}
with $q_i$ to be determined by symmetry requirements.
In Appendix~\ref{appendixB} some explicit computations for $i=0$ are presented.

This in fact turns out to be an interesting combinatorics problem. Each symmetry is associated to a distinct $q_i$, which is determined by symmetry requirements. In particular, we found the following values for $q_i$, associated to the four representations
$\mathrm{f}_i$:
\begin{gather*}
q_0 = a_1 a_2 a_3,\qquad
q_1 = a_1^{-1} a_2 a_3,\qquad
q_2 = a_1 a_2^{-1} a_3\qquad
q_3 = a_1 a_2 a_3^{-1} \nonumber
\end{gather*}
(see also the Appendix~\ref{appendixA} on the solution of the Hecke condition).

To summarize, this is the announced non trivial symmetry of the junction representation. For
the simplest case, $n=3$, we have extracted the extra non-trivial symmetry of the junction
representation, i.e.
\begin{gather*}
{\cal G}\equiv U_{q_{0}}(sl_2)\otimes U_{q_{1}}(sl_2)\otimes U_{q_{2}}(sl_2)\otimes U_{q_{3}}(sl_2),
\end{gather*}
essentially dif\/ferent from the manifest $\bigotimes_{i=1}^{3} U_{a_i}(sl_2)$ symmetry. It is important
to note that this extra symmetry is structurally much richer than the manifest one, and is enlarged almost
exponentially with respect to~$n$, as will become transparent in the subsequent section.

\subsection{Generic junction representation and exact symmetries}\label{section2.2}

We shall generalize here the investigation of the exact symmetries associated to the generic junction representation $(n>3)$.
Inspired by the $n=3$ case, and the respective form of the
representations that have been obtained there, we propose a big family
of representations below for generic values of $n$. Actually, after inspection, one observes that
the whole analysis merely reduces to a combinatorial problem.

Let us describe the generic family of representations of the quantum algebra.
Consider f\/irst the representation  ${\mathrm f}_0: U_{q_0}(sl_2) \mapsto
\mbox{End}(({\mathbb C}^{(2)})^{\otimes n})$ as a starting point
\begin{gather*}
{\mathrm f}_0(h)=(e_{11})^{\otimes n}-(e_{22})^{\otimes n}\equiv h^{(0)}_1-h^{(0)}_2,\qquad {\mathrm f}_0(e)=(e_{12})^{\otimes n}, \qquad {\mathrm f}_0(f)=(e_{21})^{\otimes n},\nonumber\\
{\mathrm f}_0(q_0^h)= \mathbb{I}+(q_0-1)h^{(0)}_1+(q_0^{-1}-1)h^{(0)}_2.
\end{gather*}
Recall that in the generic situation, $n$ constants $a_i$ are involved. The symmetry algebras exa\-mi\-ned here are
in fact associated
to all possible combinations of $a_i$'s and their inverses. More precisely, consider f\/irst the combination
$a_1 a_2\cdots a_n$, which is in fact equal to~$q_0$. By turning one of the $a_i$ to its inverse we obtain
$n$ such combinations.
Then we may turn two, three $a_i$'s etc. into their inverses, and stop turning after we reach the values
$\frac{n}{2}$ or $\frac{n-1}{2}$ (for $n$ even and odd respectively).
We stop basically because we reproduce the same algebras as we shall see below.
In fact, by repeating the procedure above we end up with a total number of combinations given~by:
\begin{gather}
M = \sum_{m=0}^{\frac{n-1}{2}}  \frac{n!}{m! (n-m)!}\qquad \mbox{if $n$ odd},\nonumber\\
M = \sum_{m=0}^{\frac{n}{2}-1} \frac{n!}{m! (n-m)!} + \frac{1}{2} \frac{n!}{\frac{n}{2}!\frac{n}{2}!}
\qquad \mbox{if $n$ even.} \label{comb}
\end{gather}
We can now introduce the following generic family of representations ${\mathrm f}_{i_1 i_2 \dots i_m}$, $m \in \{0, \ldots,  L\}$ where
$L = \frac{n}{2}$ for $n$ even and $L =\frac{n-1}{2}$ for $n$ odd (for $m=0$ we basically obtain ${\mathrm f}_0$).
The corresponding
deformation parameter is def\/ined as
\[
q_{i_1 i_2\dots i_m} = (-1)^{n+1} a_1  a_2\cdots a_{i_1}^{-1}  a_{i_1 +1} \cdots a_{i_2}^{-1}\cdots a_{i_m}^{-1}\cdots a_{n},
\]
that is each one of the constants $a_{i_j}$, $i_j \in \{1,\ldots, n\}$ is turned to its inverse.
It is thus clear that the total number of representations is given by the total number of combinations $M$ given in~\eqref{comb}.
In fact, for every $m$ we have $\frac{n!}{m! (n-m)!}$ number of representations, provided that for $m=0$ there is only one representation,
the ${\mathrm f}_0$.
Note also that
\[
q_{i_1\dots i_m} =q_{{\mathbb P}(i_1\dots i_m)},
\]
where ${\mathbb P}$ denotes all possible permutations among the elements $i_j$.
Notice that in the case of $n$
even and $m =\frac{n}{2}$
we keep only half of the combinations. For instance take $n=4$; we should only consider
$q_{34}$, $q_{24}$, $q_{23}$. It is clear that
$q_{12}= q_{34}^{-1}$,  ${\mathrm f}_{12}(e)= {\mathrm f}_{34}(f)$, ${\mathrm f}_{12}(f) = {\mathrm f}_{34}(e)$,
and similarly for the rest
(see also Appendix~\ref{appendixC} for a more detailed description of the algebra elements).
More generally, let us introduce the conjugate index:
\[
\bar i = n-i+1.
\]
It is then clear that
\[
q_{i_1\dots i_\frac{n}{2}} = q^{-1}_{\bar i_1\dots \bar i_\frac{n}{2}}.
\]

It is worth noting that some kind of duality is manifest throughout this process.
In particular, the quantum algebra with parameter $q_{i_1\dots i_m}$ is equivalent to the algebra with parameter $q^{-1}_{i_1\dots i_m}$, after interchanging $e \leftrightarrow f$ and
$h \leftrightarrow -h$. This is the reason why we stop the procedure of turning $a_i$'s to their inverses after
reaching the values $\frac{n}{2}$ or $\frac{n-1}{2}$ for $n$ even and odd respectively, as described earlier
in the text. Consequently, physical systems which may possess such underlying structures
are expected to enjoy some kind of physical equivalence that ref\/lect this duality.

A generic representation is then def\/ined along the lines described above as
\begin{gather}
{\mathrm f}_{i_1 i_2 \dots i_m}: \
 U_{q_{i_1 i_2\dots i_m}}(sl_2) \to \mbox{End}(({\mathbb C}^2)^{\otimes n}),\nonumber\\
{\mathrm f}_{i_1 i_2\dots i_m}(h) =
e_{11} \otimes \cdots e_{11} \otimes  \cdots \otimes \underbrace{e_{22}}_{i_{1}}\otimes e_{11} \cdots \otimes
\underbrace{e_{22}}_{i_{2}}\otimes  \cdots \otimes \underbrace{e_{22}}_{i_{m}}\otimes  \cdots \otimes e_{11}  \nonumber\\
\phantom{{\mathrm f}_{i_1 i_2\dots i_m}(h) =}{} -
e_{22} \otimes \cdots e_{22} \otimes  \cdots \otimes \underbrace{e_{11}}_{i_{1}}\otimes e_{22} \cdots \otimes
\underbrace{e_{11}}_{i_{2}}\otimes  \cdots \otimes \underbrace{e_{11}}_{i_{m}}\otimes  \cdots \otimes e_{22} \nonumber\\
\phantom{{\mathrm f}_{i_1 i_2\dots i_m}(h)} \equiv
h_1^{(i_1\dots i_m)} - h_2^{(i_1\dots i_m)}\label{nontri}\\
{\mathrm f}_{i_1 i_2\dots i_m}(e) = e_{12} \otimes \cdots e_{12} \otimes  \cdots \otimes \underbrace{e_{21}}_{i_{1}}\otimes
e_{12} \cdots \otimes
\underbrace{e_{21}}_{i_{2}}\otimes  \cdots \otimes \underbrace{e_{21}}_{i_{m}}\otimes  \cdots \otimes e_{12},\nonumber\\
{\mathrm f}_{i_1 i_2\dots i_m}(f) = e_{21} \otimes \cdots e_{21} \otimes  \cdots \otimes
\underbrace{e_{12}}_{i_{1}}\otimes e_{21} \cdots \otimes
\underbrace{e_{12}}_{i_{2}}\otimes  \cdots \otimes \underbrace{e_{12}}_{i_{m}}\otimes  \cdots \otimes e_{21}. \nonumber
\end{gather}
All the above quantities emerge from ${\mathrm f}_0$
by exchanging:
\begin{gather*}
e_{11} \leftrightarrow e_{22} \qquad \mbox{for} \ \ h, \qquad
e_{12} \leftrightarrow e_{21} \qquad \mbox{for} \ \  e, \ f,\nonumber
\end{gather*}
for every site  $i_j \in \{1, 2, \ldots, n\}$.

In a similar fashion the generalized representation for the $q^h$ element has the form
\[
{\mathrm f}_{q_{i_1\dots i_m}}\big(q_{i_1\dots i_m}^h\big)=\mathbb{I}+(q_{i_1\dots i_m}-1)h^{(i_1\dots i_m)}_1+
\big(q_{i_1\dots i_m}^{-1}-1\big)h^{(i_1\dots i_m)}_2.
\]
To prove that the above relations def\/ine a representation of the $U_{q_{i_1\dots i_m}}(sl_2)$ algebra is a straightforward task. One just takes into account the basic property: $e_{ij} e_{kl} = \delta_{jk} e_{il}$.

Having obtained this big family of $U_{q_{i_1\dots i_m}}(sl_2)$ representations, we conjecture that these are also
non-trivial exact symmetries of the junction representation. More precisely, our main conjecture is that:
\begin{gather}
[{\mathrm f}_{i_1\dots i_m}^{\otimes N}(\Delta^{(N)}(x)),  \Theta({\mathbb U}_l)] =0,
\qquad x \in U_{q_{i_1\dots i_m}}(sl_2). \label{symm1}
\end{gather}
This is a plausible conjecture given the results for $n=3$ and the particular structure of the mentioned representations.
Note that we have also checked the validity of our conjecture for the cases $n=4, 5$ with the use of the
algebraic program \texttt{Mathematica}.

Let us make some general qualitative comments that further enforce our claim.
Consider f\/irst the multiplication\footnote{We focus here on the ${\mathrm f}_0$
representation but our arguments may be applied for the generic family of  representations.} $\Theta(\mathbb{U}_1){\mathrm f}^{\otimes 2}_0(\Delta(h))$.
Note that $\Theta$ contains a large number of dif\/ferent terms inside, which can be grouped into powers
of $a$'s. (From now on and only for the following argument we let $a_i=a y_i$ and count in powers of~$a$.)
Actually the number of terms of a specif\/ic power $\mathcal{O}(a^m)$ is given in terms of the binomial coef\/f\/icient,
but this is not so signif\/icant at the moment. On the other hand, it is instructive to note that the terms
containing $a$'s are either~$e_{11}$ or~$e_{22}$, while those not containing $a$'s are $e_{12}$ and $e_{21}$.

Consider now the multiplication of $\Theta$ with the f\/irst term of ${\mathrm f}^{\otimes 2}_0(\Delta(h))$, that is with
\[
(e_{11})^{\otimes n}\otimes \mathbb{I}^{\otimes n}.
\]
Recall that $\Theta$ is naturally ``broken'' into two parts, each one living on $(\mathbb{C}^2)^{\otimes n}$.
We see that in order for the product of $\Theta$ with the above term to
be dif\/ferent from zero, the f\/irst part of $\Theta$ should only contain the basis vectors $e_{11}$
and $e_{21}$, in every possible combination. Otherwise it just gives zero. Consequently, the problem is
reduced into a combinatorics problem. However, for the case of ${\mathrm f}^{\otimes 2}_0(\Delta(h))$
the situation is easy to handle\footnote{For the co-products of the other two elements
the situation is more dif\/f\/icult to handle, since the products with $\Theta$ do not vanish
identically. Hence the various conditions relating $a_i$'s must be taken into account, which
makes the presentation of a general argument a dif\/f\/icult task.}, and we see that the product
$\Theta(\mathbb{U}_1){\mathrm f}^{\otimes 2}_0(\Delta(h))$ vanishes identically. The reason is the following.

We observe that from the multiplication of $\Theta$ with the f\/irst term of ${\mathrm f}^{\otimes 2}_0(\Delta(h))$ only some specif\/ic
terms survive. However, these are exactly the same terms that survive the multiplication of $\Theta$
with the last term of the co-product, namely with
\[
\mathbb{I}^{\otimes n}\otimes(e_{22})^{\otimes n}.
\]
And since those two terms appear with a dif\/ferent sign, they cancel out. The same happens with
the other two terms. The same argument is also valid for the second term of the commutator, that is
for the multiplication ${\mathrm f}^{\otimes 2}_0(\Delta(h))\Theta$, which is identically zero. Hence, we have shown~-- using a
somehow heuristic argument~-- that
\[
\big[\Theta(\mathbb{U}_1),{\mathrm f}_0^{\otimes 2}(\Delta(h))\big]=0.
\]
Note that this result is valid for all the representations that belong to the
generic family that we have considered so far, due to the highly symmetric nature of ${\mathrm f}^{\otimes 2}_{i_1\dots i_m}(\Delta(h))$.
It should be stressed that although the above argument is not a rigorous proof, we expect
that it holds for the whole family of representations. Similar arguments, albeit more complicated
in their technicalities, can be presented for the other two elements of the quantum algebra.

Finally, let us brief\/ly discuss for completeness the existence of trivial symmetries associated to the junction representation. Def\/ine:
\begin{gather}
\pi_i(x) =  {\mathbb I}\otimes \cdots \otimes \cdots \underbrace{\pi_{a_i}(x)}_{1^{(i)}}\otimes \cdots
\otimes {\mathbb I}, \qquad x \in U_{a_i}(sl_2). \label{tri}
\end{gather}
It is straightforward then to check that the junction representation commutes with the actions def\/ined above.
Each one of the representations $\rho_{a_i}$ acts non trivially on the spaces ${\mathbb C}^2_{l^{(i)}}
\otimes {\mathbb C}_{(l+1)^{(i)}}^2$ separately, that is
\[
\big[\rho_{a_i}(U_{l^{(i)}}),  \pi_j^{\otimes N}(\Delta^{(N)}(x))\big] =0, \qquad i, j \in \{1, \ldots ,n\},
\]
which immediately leads
to
\begin{gather}
\big [\Theta({\mathbb U}_l),  \pi_i^{\otimes N}(\Delta^{(N)}(x)) \big ] =0, \qquad x \in U_{a_i}(sl_2), \label{basic2}
\end{gather}
exposing the manifest symmetry of a generic junction representation
\begin{gather}
{\cal G}_0 \equiv U_{a_1}(sl_2)\otimes U_{a_2}(sl_2)\otimes \cdots\otimes U_{a_n}(sl_2). \label{basic4}
\end{gather}

To summarize, we have been able to explicitly prove the existence of non-trivial exact symmetries
\eqref{rep}, \eqref{symm0} associated to the junction representation for $n=3$
(see Appendix~\ref{appendixB}). We have conjectured the generic form of non-trivial symmetries of the junction
representation \eqref{nontri} $\forall\, n$, and we have checked the validity of our conjecture
\eqref{symm1} with \texttt{Mathematica} for $n=4,  5$. The existence of trivial symmetries~\eqref{tri}
is an immediate consequence of the structure of the junction representation, therefore the proof of
\eqref{basic2}, \eqref{basic4} is straightforward $\forall\, n$.

\section[Boundary Temperley-Lieb algebra and representations]{Boundary Temperley--Lieb algebra and representations}\label{section3}

We shall discuss in this section extensions of the junction representation in the case of the boundary Temperley--Lieb algebra.
The boundary Temperley--Lieb $B_N(q, Q)$ algebra is def\/ined by generators ${\mathbb U}_i$ obeying~\eqref{tl}
and ${\mathbb U}_0$ satisfying
\begin{gather*}
{\mathbb U}_0^2 = \delta_0  {\mathbb U}_0,\qquad
{\mathbb U}_1  {\mathbb U}_0  {\mathbb U}_1 = \kappa  {\mathbb U}_1, \qquad
[{\mathbb U}_0,  {\mathbb U}_i] =0, \qquad i>1,\nonumber
\end{gather*}
where $\delta_0=- (Q+Q^{-1})$, $\kappa = qQ^{-1}+q^{-1}Q$, although there is a reparametrization freedom that one can appropriately use (see e.g. \cite{doikoumartin}).

The well known XXZ representation of the boundary Temperley--Lieb algebra is given next.
Let $U$ be the matrix def\/ined in \eqref{U_XXZ} and
\begin{gather}
U_0 = -Q^{-1}e_{11} - Qe_{22} +e_{12} +e_{21}. \label{tr1}
\end{gather}
Then the following is a representation of the boundary TL algebra,
\begin{gather}
\rho_{q,Q}: \  B_N(q,Q) \mapsto \mbox{End}\big(({\mathbb C}^{2})^{\otimes N}\big),\nonumber\\
\rho_{q,Q}({\mathbb U}_i) = {\mathbb I} \otimes \cdots \otimes {\mathbb I} \otimes \underbrace{U}_{i,\, i+1}
\otimes {\mathbb I}\otimes
\cdots \otimes {\mathbb I}, \qquad
\rho_{q,Q} ({\mathbb U}_0) = U_0 \otimes {\mathbb I} \otimes \cdots \otimes {\mathbb I}. \label{tr2}
\end{gather}
The obvious junction representation for the ``boundary'' element for generic $n$
is then the following
\begin{gather*}
\Theta({\mathbb U}_0) = \prod_{i=1}^n\rho_{a_i, Q_i}({\mathbb U}_0),
\end{gather*}
where each one of the XXZ representations act on the spaces ${\mathbb C}^2_{1^{(1)}}$, ${\mathbb C}^2_{1^{(2)}}$,  $\ldots$, ${\mathbb C}^2_{1^{(n)}}$;
i.e.
\[
\rho_{a_i, Q_i}({\mathbb U}_0) = {\mathbb I} \otimes \cdots \otimes \underbrace{U_0}_{1^{(i)}}
\otimes \cdots \otimes {\mathbb I}.
\]
It is straightforward to check, due to the fact that spaces ${\mathbb C}^2_{l^{(i)}}$ commute among each other that
$\Theta({\mathbb U}_i)$ and $\Theta({\mathbb U}_0)$ do satisfy the algebraic relations of the boundary TL algebra,
thus they provide a representation of the algebra.

However, our objective here is to search
for non trivial boundary elements that mix essentially the spaces exactly as the non trivial symmetries mix the various spaces
${\mathbb C}^2_{1^{(i)}}$. Consider the following matrices
\[
M_{i_1\dots i_m} = -Q^{-1}h_1^{(i_1\dots i_m)} -Qh_2^{(i_1\dots i_m)} + {\mathrm f}_{i_1\dots i_m}(e)+
{\mathrm f}_{i_1\dots i_m}(f),
\]
and recall that $h_{1,2}^{(i_1\dots i_m)}$ and ${\mathrm f}_{i_1\dots i_m}(e,  f)$
are def\/ined in Section~\ref{section2.2}. The indices take values $i_j \in \{1, \ldots , n\}$ as also
described in Section~\ref{section2.2}.

Let us focus again on the $n=3$ case, where analytic proofs of our claims are available (for higher $n$ analytic proofs are highly complicated, and only checks with the help of algebraic packages can be performed).
First, introduce the
following matrices corresponding to the four~$\textrm{f}_i$ representations of the $n=3$ case described in
Section~\ref{section2.1},
\begin{gather*}
M_0 = -Q^{-1} e_{11} \otimes e_{11} \otimes e_{11} - Q e_{22} \otimes e_{22} \otimes e_{22} +
e_{12} \otimes e_{12} \otimes e_{12} +e_{21} \otimes e_{21} \otimes e_{21}, \nonumber\\
M_1 = -Q^{-1} e_{22} \otimes e_{11} \otimes e_{11} - Q e_{11} \otimes e_{22} \otimes e_{22} + e_{21} \otimes e_{12} \otimes e_{12}
+e_{12} \otimes e_{21} \otimes e_{21},\nonumber\\
M_2 = -Q^{-1} e_{11} \otimes e_{22} \otimes e_{11} -Q e_{22} \otimes e_{11} \otimes e_{22} + e_{12} \otimes e_{21} \otimes e_{12} +
e_{21} \otimes e_{12} \otimes e_{21}, \nonumber\\
M_3 = -Q^{-1} e_{11} \otimes e_{11} \otimes e_{22} -Q e_{22} \otimes e_{22} \otimes e_{11} + e_{12} \otimes e_{12} \otimes e_{21}
+e_{21} \otimes e_{21} \otimes e_{12}.\nonumber
\end{gather*}
The matrices $M_i $ act on ${\mathbb C}^2_{1^{(1)}}\otimes {\mathbb C}^2_{1^{(2)}} \otimes {\mathbb C}^2_{1^{(3)}}$. Also set
\[
{\cal M}_i = M_i \otimes {\mathbb I}\otimes \cdots \otimes{\mathbb I},
\]
where now ${\mathbb I}$ acts on $({\mathbb C}^2)^{\otimes 3}$. It can be shown then that
\[
\Theta({\mathbb U}_1)  {\cal M}_i \Theta({\mathbb U}_1) = \kappa \Theta({\mathbb U}_1),
\qquad i=0,1,2,3,
\]
and this suf\/f\/ices to prove that this is a representation of the boundary Temperley--Lieb algebra.
Indeed we def\/ine the representation $\Theta_i:  B_N(q,Q) \mapsto
\mbox{End}({\mathbb C}^2 \otimes {\mathbb C}^2 \otimes {\mathbb C}^2 )^{\otimes N}$,  $i \in \{0, 1, 2, 3\}$ such that:
\[
\Theta_i({\mathbb U}_l) = \Theta({\mathbb U}_l),\qquad  l \in \{1,  2, \ldots, N-1\},
\qquad \Theta_i({\mathbb U}_0) = {\cal M}_i.
\]

We can proceed then and state the following generic conjecture $\forall\, n>3$. We conjecture
that
$\Theta_{i_1\dots i_m}$ such that
\begin{gather*}
\Theta_{i_1\dots i_m}({\mathbb U}_l) = \Theta({\mathbb U}_l), \qquad l \in \{1, \ldots, N-1\},
\qquad \Theta_{i_1\dots i_m}({\mathbb U}_0) = {\cal M}_{i_1\dots i_m}, \nonumber\\
{\cal M}_{i_1\dots i_m} = M_{i_1\dots i_m}\otimes {\mathbb I} \otimes \cdots \otimes {\mathbb I}
\end{gather*}
is a representation of the boundary Temperley--Lieb algebra. To prove this it is suf\/f\/icient to show that
\[
\Theta({\mathbb U}_1)  {\cal M}_{i_1\dots i_m}  \Theta({\mathbb U}_1) = \kappa \Theta({\mathbb U}_1).
\]
As before, we have checked the validity of the latter conjecture by performing an analytical
computation for $n=3$ and by using \texttt{Mathematica} for the cases $n=4,  5$. Therefore, we have strong indications that it is valid for generic $n$. The analytic proof for $n=3$ is omitted here for brevity, but it is straightforward, although technically demanding, and is based on the explicit form of $\Theta({\mathbb U}_1)$, ${\cal M}_{i_1\dots i_m}$ and the property: $e_{ij} e_{kl} = \delta_{jk} e_{il}$

As mentioned above, the structure of these boundary elements is inspired by the form of the non trivial
symmetries studied in the previous section. It will be transparent in the subsequent section that they break the
respective symmetries then, along the lines described in~\cite{doikoumartin, doikouhecke, doikoutwin}.

\subsection{Residual symmetries}\label{section3.1}

In this section we shall examine how each one of the non-trivial boundary elements ${\cal M}_{i_1\dots i_m}$ breaks the original symmetry of the Temperley--Lieb algebra. This will be particularly useful when studying the corresponding quantum spin chain symmetry.

Apart from the manifest symmetry of the algebra, we have exposed in the previous
section a~$U_{q_0}(sl_2)\otimes U_{q_1}(sl_2)\otimes U_{q_2}(sl_2)
\otimes U_{q_3}(sl_2)$ symmetry for the junction representation. It is
straightforward to check that each one of the non-trivial boundary elements
breaks only part of this symmetry, while preserving the rest of it. In
particular, it is shown that
\[
[{\mathrm f}_i(x),  M_j]=0, \qquad x\in U_{q_i}(sl_2), \qquad i\neq j.
\]
The above condition is not accidental at all. In fact, it is somehow expected,
given that the structure of those boundary elements is strictly related to the respective
non-trivial symmetries of the algebra. However, the case where $i=j$ requires
particular attention.

Let us f\/irst consider the following combination of generators of the quantum algebra
\linebreak $U_{q_{i_1\dots i_m}}(sl_2)$ (see also e.g.~\cite{doikoutwin, mene, dema})
\begin{gather*}
\mathcal{Q}_{i_1\dots i_m}=q_{i_1\dots i_m}^{-\frac{1}{2}}q_{i_1\dots i_m}^\frac{h}{2}
 e+q_{i_1\dots i_m}^{\frac{1}{2}}q_{i_1\dots i_m}^{-\frac{h}{2}} f+
x_{i_1\dots i_m}q_{i_1\dots i_m}^{h}-x_{i_1\dots i_m}\mathbb{I},
\end{gather*}
where $x_{i_1\dots i_m}$ are constants to be determined later on. Focus again on the simplest non-trivial situation $n=3$.
Using then
the four ${\mathrm f}_i$ representations at hand one may compute the respective
charges, which have the following form
\begin{gather*}
{\mathrm f}_0(\mathcal{Q}_{0}) = (-x_{0}+q_0x_{0})e_{11}\otimes e_{11}\otimes e_{11}+e_{12}\otimes e_{12}\otimes e_{12}\nonumber\\
\phantom{{\mathrm f}_0(\mathcal{Q}_{0}) =}{} +e_{21}\otimes e_{21}\otimes e_{21}+(-x_{0}+q_0^{-1}x_{0})e_{22}\otimes e_{22}\otimes e_{22},\nonumber\\
{\mathrm f}_1(\mathcal{Q}_{1}) = (-x_{1}+q_1x_{1})e_{22}\otimes e_{11}\otimes e_{11}+e_{21}\otimes e_{12}\otimes e_{12}\nonumber\\
\phantom{{\mathrm f}_1(\mathcal{Q}_{1}) =}{}
+e_{12}\otimes e_{21}\otimes e_{21}+(-x_{1}+q_1^{-1} x_{1})e_{11} \otimes e_{22}\otimes e_{22},\nonumber\\
{\mathrm f}_2(\mathcal{Q}_{2}) = (-x_{2}+q_2x_{2})e_{11}\otimes e_{22}\otimes e_{11}+e_{12}\otimes e_{21}\otimes e_{12}\nonumber\\
\phantom{{\mathrm f}_2(\mathcal{Q}_{2}) =}{}
+e_{21}\otimes e_{12}\otimes e_{21}+(-x_{2}+q_2^{-1}x_{2})e_{22}\otimes e_{11}\otimes e_{22},\nonumber\\
{\mathrm f}_3(\mathcal{Q}_{3}) = (-x_{3}+q_3x_{3})e_{11}\otimes e_{11}\otimes e_{22}+e_{12}\otimes e_{12}\otimes e_{21}\nonumber\\
\phantom{{\mathrm f}_3(\mathcal{Q}_{3}) =}{}
+e_{21}\otimes e_{21}\otimes e_{12}+(-x_{3}+q_3^{-1}x_{3})e_{22}\otimes e_{22}\otimes e_{11}.\nonumber
\end{gather*}
This charge commutes with all the irrespective boundary elements, that is
\[
[{\mathrm f}_i(\mathcal{Q}_{i}),  M_j]=0  \qquad \textrm{for} \ \  i\neq j,
\]
but is also found to commute with the relevant boundary element
\[
[{\mathrm f}_i(\mathcal{Q}_{i}),  M_i]=0,
\]
provided that the constants $x_{i}$ satisfy
\[
x_{i} = \frac{Q-Q^{-1}}{q_i-q_i^{-1}}.
\]
To conclude, the symmetry we extract here for each representation $\Theta_i$
of the boundary Tem\-per\-ley--Lieb algebra is:
\[
\big[\Theta_i({\mathbb U}_l),  {\mathrm f}^{\otimes N}_i(\Delta^{(N)}({\cal Q}_i))\otimes_{i \neq j}
{\mathrm f}_j^{\otimes N}(\Delta^{(N)}(x))\big] =0,
\qquad x \in U_{q_j}(sl_2).
\]
We may generalize the statement -- as a conjecture -- on the boundary symmetry $\forall \, n$, i.e., we conjecture that
\begin{gather}
[\Theta_{i_1\dots i_m}({\mathbb U}_l), {\mathrm f}^{\otimes N}_{i_1\dots i_m}(\Delta^{(N)}({\cal Q}_{i_1\dots i_m}))\otimes_{\{i_1,\dots ,i_m\} \neq \{j_1,\dots ,j_m\}}
{\mathrm f}_{j_1\dots j_m}^{\otimes N}(\Delta^{(N)}(x))]=0, \nonumber\\
\qquad x \in U_{q_{j_1\dots j_m}}(sl_2). \label{symm2}
\end{gather}
It is clear that the trivial representation def\/ined in \eqref{tr1}, \eqref{tr2} enjoys the symmetry (see also~\cite{doikoutwin}):
\begin{gather*}
\big[\Theta_i({\mathbb U}_l), \pi^{\otimes N}_i(\Delta^{(N)}({\cal Q}_i))\otimes_{i \neq j}
\pi_j^{\otimes N}(\Delta^{(N)}(x))\big] =0,
\qquad x \in U_{a_j}(sl_2), \qquad i,  j \in \{1, \dots,n\},
\end{gather*}
where we def\/ine
\[
\mathcal{Q}_{i}=a_{i}^{-\frac{1}{2}}a_{i}^\frac{h}{2} e+a_{i}^{\frac{1}{2}}a_{i}^{-\frac{h}{2}} f+
x_{i}a_{i}^h-x_{i}\mathbb{I}, \qquad \! \! i \in \{1, \ldots ,n\},
\]
and
\[
x_i = \frac{Q - Q^{-1}}{a_i - a_i^{-1}}.
\]
For a more detailed discussion on boundary symmetries we refer the interested reader to \cite{doikoumartin, doikoutwin}.

\section{The quantum spin chain}\label{section4}

We shall now brief\/ly discuss the corresponding quantum spin chain, enjoying the symmetries described above.
Recall the $\check R$ matrix is a solution of the braid Yang--Baxter equation
\[
\check R_{12}(\lambda_1 -\lambda_2)  \check R_{23}(\lambda_1) \check R_{12}(\lambda_2) =
\check R_{23}(\lambda_2)  \check R_{12}(\lambda_1)  \check R_{23}(\lambda_1 -\lambda_2).
\]
The $\check R$ matrix associated with representations of the Temperley--Lieb algebra may be expressed as~\cite{jimbo}
\[
\check R_{i i+1}(\lambda) = \sinh(\lambda + i\mu) + \sinh \lambda  \rho({\mathbb U}_i),
\]
for any representation $\rho$ of the TL algebra $T_N(q)$, and where $q =e^{i\mu}$.

Def\/ine the $R$ matrix as: $R(\lambda) = {\cal P}  \check R(\lambda)$. The $R$ matrix
associated with the junction representation satisf\/ies unitarity and
crossing  symmetry i.e.
\[
R_{ij}(\lambda) R_{ji}(-\lambda) \propto {\mathbb I},\qquad R_{ij}(\lambda)
\propto {\mathbb V}_i  R^{t_j}_{ij}(-\lambda -i )  {\mathbb V}_i,
\]
where we def\/ine in general
\[
{\mathbb V} =  \otimes_{i=1}^n\left( \begin{array}{cc}
	                         & a_{i}^{-{1\over 2} } \\
				   a_{i}^{{1\over 2}}  &
\end{array} \right),
\]
and for $n=3$ in particular we have:
\begin{gather}
{\mathbb V} =  \left( \begin{array}{cc}
	                         & a_{1}^{-{1\over 2} } \\
				   a_{1}^{{1\over 2}}  &
\end{array} \right)  \otimes \left( \begin{array}{cc}
	                         & a_{2}^{-{1\over 2} } \\
				   a_{2}^{{1\over 2} }  &
\end{array} \right)\otimes \left( \begin{array}{cc}
	                         & a_{3}^{-{1\over 2} } \\
				   a_{3}^{{1\over 2} }  &
\end{array} \right) \nonumber\\
\phantom{{\mathbb V}}{} =  \left( \begin{array}{cccccccc}
	                 &&&&&&&q_0^{-{1\over 2}} \\
				     &&&&&&q_3^{-{1\over 2}}&\\
&&&&&q_2^{-{1\over 2}}&&\\
&&&&q_1^{{1\over 2}}&&&\\
&&&q_1^{-{1\over 2}}&&&&\\
&&q_2^{{1\over 2}}&&&&&\\
&q_3^{{1\over 2}}&&&&&&\\
q_0^{{1\over 2}}&&&&&&&\\
\end{array} \right).\nonumber
\end{gather}
Recall also the identif\/ications of $q_i$'s, entailed by the exact symmetries of the
representation.

Then one may construct the transfer matrix of an open spin chain with generic integrable
boundary conditions \cite{sklyanin}:
\begin{gather*}
t(\lambda)= {\rm tr}_0\{M_0  K_0^+  {\mathbb T} (\lambda) \},
\end{gather*}
where
\begin{gather*}
{\mathbb T}= T_0(\lambda)  K_0(\lambda)  T_0^{-1}(\lambda), \qquad T(\lambda) =
R_{0N}(\lambda)\cdots R_{01}(\lambda).
\end{gather*}
The $K$ matrix satisf\/ies the ref\/lection equation
\[
R_{12}(\lambda_1 -\lambda_2) K_1(\lambda_1) R_{21}(\lambda_1 +\lambda_2)K_2(\lambda_2) = K_2(\lambda_2) R_{12}(\lambda_1+\lambda_2)
K_1(\lambda_1) R_{21}(\lambda_1-\lambda_2).
\]
We shall focus here on the case of trivial left boundary, that is $K^+ \propto {\mathbb I}$.
Also $M$ is def\/ined as
\[
M = {\mathbb V}^t  {\mathbb V},
\]
and for $n=3$ in particular it becomes:
\[
M = \left( \begin{array}{cccccccc}
	                 q_0&&&&&&& \\
				     &q_3&&&&&&\\
&&q_2&&&&&\\
&&&q_1^{-1}&&&&\\
&&&&q_1&&&\\
&&&&&q_2^{-1}&&\\
&&&&&&q_3^{-1}&\\
&&&&&&&q_0^{-1}\\
\end{array} \right).
\]
It is known that solutions of the ref\/lection equation may be expressed in terms of representations of the boundary Temperley--Lieb algebra as (see also e.g. \cite{levymartin, doikoumartin}):
\begin{gather*}
K(\lambda) =x(\lambda) + y(\lambda) \rho({\mathbb U}_0), \nonumber\\
x(\lambda)=-\delta_0\cosh(2\lambda +i\mu) - \kappa \cosh(2\lambda) -\cosh(2i\zeta),\nonumber\\
y(\lambda) = 2 \sinh(2\lambda) \sinh(i\mu).\nonumber
\end{gather*}

Let us f\/irst consider the case of a trivial right boundary i.e.~$K \propto {\mathbb I}$. Then as was proven in
Section~\ref{section2.1} that the junction representation enjoys the $\otimes_{i=0}^3 U_{q_i}(sl_2)$ symmetry and consequently
\begin{gather*}
\big[\check R,  {\mathrm f}_i^{\otimes 2}(\Delta^{\otimes 2}(x))\big] = 0 \quad
  \Rightarrow \quad {\mathrm f}_i^{\otimes 2}(\Delta^{'\otimes 2}(x))  R(\lambda) = R(\lambda)  {\mathrm f}_i^{\otimes 2}
(\Delta^{\otimes 2}(x)),\qquad x \in U_{q_i}(sl_2),
\end{gather*}
where $\Delta' = {\cal P}  \Delta  {\cal P}$ and
where $R$ is the junction $R$ matrix. From the relations above it is straightforward to show that:
\[
{\mathrm f}_i^{\otimes (N+1)}(\Delta^{'\otimes (N+1)}(x)) T(\lambda) =T(\lambda)
\Delta^{\otimes (N+1)}({\mathrm f}_i^{\otimes (N+1)}(x)),
\]
which leads, for trivial boundary conditions to
\begin{gather*}
{\mathrm f}_i^{\otimes (N+1)}(\Delta^{'\otimes (N+1)}(x))  {\mathbb T}(\lambda) = {\mathbb T}(\lambda)
{\mathrm f}_i^{\otimes (N+1)}(\Delta^{'\otimes (N+1)}(x)) \\
\qquad \Rightarrow \quad
[t(\lambda),  {\mathrm f}_i^{\otimes (N)}(\Delta^{\otimes (N)}(x)) ]=0.\nonumber
\end{gather*}
Hence, the transfer matrix with trivial boundary conditions
enjoys the  $\otimes_{i=0}^3 U_{q_i}(sl_2)$ symmetry (similarly it may be shown that the transfer matrix enjoys
the obvious symmetry $\otimes_{i=1}^3U_{a_i}(sl_2)$, see also \cite{doikouhecke, doikoutwin} for further details).
The statement may be generalized for any ${\mathrm f}_{i_1\dots i_m}$ representation $\forall \, n$ provided that relations \eqref{symm1} are valid:
\[
\big[t(\lambda), {\mathrm f}_{i_1\dots i_m}^{\otimes (N)}(\Delta^{\otimes (N)}(x)) \big]=0, \qquad x
\in U_{q_{i_1\dots i_m}}(sl_2).
\]

We shall focus here, since it is more interesting, on the breaking of the non trivial symmetries expressed
via the representations ${\mathrm f}_i$.
The existence of a non-trivial ``boundary'' may couple the various $V_{l^{(i)}}$ spaces. In this case various choices of representations of the boundary Temperley--Lieb algebra lead to consistent breaking of the original symmetry in dif\/ferent ways. Particularly, focus f\/irst on the $n=3$ case and def\/ine
\[
K^{(i)} = x(\lambda) + y(\lambda) M_i \qquad \mbox{and}\qquad {\mathbb T}^{(i)} = T(\lambda)  K^{(i)}
(\lambda)  T^{-1}(-\lambda).
\]
It is clear bearing also in mind the results of Section~\ref{section3} that
\begin{gather*}
\big[K^{(i)}(\lambda),  {\mathrm f}_i({\cal Q}_i) \otimes_{j\neq i} {\mathrm f}_j(x)\big] =0\\
\qquad {} \Rightarrow \quad
{\mathbb T}^{(i)}(\lambda)  {\mathrm f}^{\otimes (N+1)}_i (\Delta^{'(N+1)}({\cal Q}_i)) \otimes_{j\neq i}
{\mathrm f}_j^{\otimes (N+1)}(\Delta^{'(N+1)}(x))\nonumber\\
\qquad {} ={\mathrm f}^{\otimes (N+1)}_i
(\Delta^{'(N+1)}({\cal Q}_i)) \otimes_{j\neq i}
{\mathrm f}_j^{\otimes (N+1)}(\Delta^{'(N+1)}(x))  {\mathbb T}^{(i)}(\lambda),
\end{gather*}
and it is then shown that
\[
\big[t^{(i)}(\lambda),  {\mathrm f}^{\otimes N}_i(\Delta^{(N)}({\cal Q}_i)) \otimes_{j \neq i} {\mathrm f}^{\otimes N}_j
(\Delta^{(N)}(x))\big] =0, \qquad x \in U_{q_j}(sl_2).
\]
With this statement we basically conclude our study on the symmetries of the junction spin chain with trivial and non-trivial
integrable boundary conditions.
Again the boundary symmetry statement can be generalized for any representation ${\mathrm f}_{i_1\dots i_m}$, $\forall \, n$ provided that
relations \eqref{symm1}, \eqref{symm2} hold. Indeed, set
\[
K^{(i_1\dots i_m)}= x(\lambda) + y(\lambda) M_{i_1\dots i_m}, \qquad t^{(i_1\dots i_m)}(\lambda)
={\rm  tr}\{M T(\lambda) K^{(i_1\dots i_m)}(\lambda)  T^{-1}(-\lambda) \}.
\]
Then it is straightforward to show, provided that \eqref{symm1}, \eqref{symm2} are valid:
\begin{gather*}
\big[t^{(i_1\dots i_m)}(\lambda),  {\mathrm f}^{\otimes N}_{i_1\dots i_m}(\Delta^{(N)}({\cal Q}_{i_1\dots i_m})) \otimes_{\{j_1,\dots ,j_m\} \neq \{i_1,\dots i_m\}} {\mathrm f}^{\otimes N}_{j_1\dots j_m}
(\Delta^{(N)}(x))\big] =0,\nonumber\\
\qquad x \in U_{q_{j_1\dots j_m}}(sl_2).\nonumber
\end{gather*}

Since the junction representation is a representation of the (boundary) TL algebra as is the XXZ one expects,
according to the general discussion in \cite{doikoumartin}, that the spectra of the models will by identical, up to multiplicities, (see universal expressions
for Hamiltonians in \cite{doikoumartin} and generic expressions of open transfer matrices
in terms of elements of the (boundary) Temperley--Lieb algebra in \cite{doikoutwin}). Note that the results of the present section are explicitly proven provided that the results, and conjectures of Sections~\ref{section2.1}, \ref{section2.2}, \ref{section3} and \ref{section3.1} are valid.

\section{Discussion}\label{section5}

We have been able to identify a novel family of representations of the Temperley--Lieb
algebra, the so called junction
representations, which involve n copies of the familiar XXZ
representation.
Interestingly enough, it turns out that the number of exact symmetries associated to the junction
representation is drastically increased with the number $n$. The existence of the extended junction
representation in the case of the boundary Temperley--Lieb algebra is examined, and particular non-trivial
``boundary'' elements are identif\/ied. The results on the boundary case are non-trivial and analytic proof is
provided in the case where $n=3$, whereas in the general case a conjecture on the form of the representation of
the boundary element is stated, and is checked with Mathematica for $n=4,\, 5$.
The associated residual symmetries in this case are also extracted.

More precisely, regarding the study of the associated symmetries: we explicitly show in Section~\ref{section2.1} that in
addition to the expected symmetries particular non-trivial algebra realizations commute with the junction
representation, whereas in Section~\ref{section3} we state a generic conjecture on the existence of a big number of
non-trivial symmetry algebras, which we support qualitatively. We check however the validity of our conjecture
with \texttt{Mathematica} up to $n=5$. Similar results obtained in Section~\ref{section3.1} concerning the associated boundary
symmetries.

What is highly non-trivial in the present investigation in the existence of a large number of non-trivial exact
symmetry algebras. It is important to note that for the moment there is no-rigorous proof that these consist the
full symmetry algebra associated to the junction representation. Nevertheless, it would be very interesting if
one could f\/ind an extra symmetry that does not fall to the categories discussed in the present study. Moreover,
the obvious number of associated symmetries does not seem to coincide with the number of the obtained non-trivial
symmetries, and this is also an intriguing issue, which is still open even in the simplest case $n=2$, the
``asymmetric twin'' representation~\cite{doikoumartin}. We hope to address these signif\/icant issues in future
investigations.

Using this particular representation we constructed the corresponding novel physical quantum system, that is the ``junction'' quantum spin chain, and examined the associated exact symmetries. Within this spirit an interesting problem to pursue is the study of the aforementioned representations, and symmetries in the context of other statistical physics, such as face or loop models.
An even more intriguing question is the possible existence of some kind of continuum limit of the junction spin chain, which also enjoys the big number of exact symmetries identif\/ied here. More precisely, universal local Hamiltonians in terms of the elements of the (boundary) Temperley--Lieb algebras are available (see e.g.~\cite{doikoumartin, doikouhecke}), thus one may try to obtain the so called long wave length continuum limit (see e.g.~\cite{fradkin}) of such expressions, that is obtain classical continuum Hamiltonians.
We shall fully address these and related issues in forthcoming publications.

\appendix
\section{Parametrization and solution of the Hecke condition}\label{appendixA}

In this appendix we present how one obtains solutions of the quadratic constraint of the
Temperley--Lieb algebra (Hecke condition)
\[
{\mathbb U}^2 = -\big(q+q^{-1}\big)  {\mathbb U}.
\]
We focus f\/irst for simplicity on the case $n=3$. A detailed study for the generic solution
will be presented elsewhere.

By taking into account the form of the junction representation, the Hecke condition is
written as
\begin{gather*}
a_1a_2a_3 +a_1^{-1} a_2^{-1}a_3^{-1} +a_1 a_{2} a_3^{-1}+ a_1 a_2^{-1}a_3+ a_1^{-1} a_2 a_3  \\
\qquad{}+ a_1^{-1} a_2^{-1} a_3
+a_1^{-1} a_2 a_3^{-1} + a_1 a_2^{-1} a_3^{-1} = q +q^{-1}.
\end{gather*}
It is now convenient to parametrize $a_i$ as follows:
\[
a_i = z_i q^{1\over 3}, \qquad i\in \{1,  2,  3\}
\]
In addition, if we set
\[
a_1a_2 a_3 = q \quad \Rightarrow \quad a_1^{-1}a_2^{-1} a_3^{-1} = q^{-1},
\]
then the rest terms sum up to give zero. Hence we are left with three equations to solve in order to identify the three unknown quantities $z_i$, which read
\begin{gather}
z_1z_2 z_3 = 1,\qquad
z_1 z_{2} z_3^{-1}+ z_1 z_2^{-1}z_3 + z_1^{-1} z_2 z_3 =0\nonumber\\
z_1^{-1} z_2^{-1} z_3 +z_1^{-1} z_2 z_3^{-1} + z_1 z_2^{-1} z_3^{-1}=0. \label{system}
\end{gather}
Solving the equations above one can explicitly f\/ix the values of~$z_i$, for the particular parametrization of~$a_i$.
Of course one may choose a dif\/ferent parametrization, but in any case these should be equivalent. A simple solution of the above system of equations is the following:
\[
z_1= e^{{i\pi \over 3}}, \qquad z_2 =e^{{2i\pi \over 3}} ,\qquad z_3= e^{i\pi }.
\]
It is straightforward to check that the above solution satisf\/ies the system~(\ref{system}).

Let us now generalize our comments for every $n>3$. We start from the generic polynomial condition
\[
(-1)^n \prod_{j=1}^n \big(a_j + a_j^{-1}\big) = -\big(q+q^{-1}\big),
\]
and expanding the expressions above we take
\[
a_1 a_2 \cdots a_n + \sum_{i} a_1 \cdots a_{i}^{-1}\cdots a_n +
\sum_{i_1, i_2} a_1 \cdots a_{i_1}^{-1}\cdots a_{i_2}^{-1}\cdots a_n + \cdots = (-1)^{n-1}\big(q+q^{-1}\big).
\]
We parameterize, as in $n=3$ case, by setting
\begin{gather}
q = (-1)^{n-1} a_1 a_2 \cdots a_n, \qquad a_j = z_j  q^{{1\over n}}. \label{one}
\end{gather}
Our task now is to identify the constants $z_i$. By also requiring, in addition to~\eqref{one}, the factor in front of
every power $q^{{m\over n}}$ ($-n< m <n$) to disappear we end up with a set of
$n$ equations. More specif\/ically, we end up with a system of $n$ unknown quantities, the
$z_j$'s and $n$ equations, so that in principle the system can be solved and $z_j$'s
can be determined.

\section[Proof for $n=3$]{Proof for $\boldsymbol{n=3}$}\label{appendixB}

As it has been already stated, we have checked the validity of our conjecture for the cases
$n=4,\,5$ by using \texttt{Mathematica}. However, we were able to perform analytic calculations
for the $n=3$ case. In this appendix, we describe part of our analytic computation.

It is convenient for what follows to write down the explicit expression of the each one of the XXZ representations for $n=3$:
i.e.
\begin{gather*}
\rho_{a_1}({\mathbb U}_{l^{(1)}}) = \cdots{\mathbb I} \otimes \Bigg(\sum_{a \neq b}
e_{ab} \otimes {\mathbb I} \otimes {\mathbb I}\otimes e_{ba} \otimes {\mathbb I}\otimes {\mathbb I} \nonumber\\
\phantom{\rho_{a_1}({\mathbb U}_{l^{(1)}}) =}{} - \sum_{a\neq b}
a_1^{-{\rm sgn}(a-b)} e_{aa}
\otimes {\mathbb I} \otimes {\mathbb I}\otimes  e_{bb} \otimes  {\mathbb I}\otimes {\mathbb I}  \Bigg)\otimes {\mathbb I}\cdots, \nonumber\\
\rho_{a_2}({\mathbb U}_{l^{(2)}}) = \cdots{\mathbb I} \otimes \Bigg(\sum_{a \neq b}
{\mathbb I}  \otimes e_{ab} \otimes {\mathbb I}\otimes {\mathbb I}  \otimes e_{ba}\otimes {\mathbb I}  \nonumber\\
\phantom{\rho_{a_2}({\mathbb U}_{l^{(2)}}) =}{}
- \sum_{a\neq b}
a_2^{-{\rm sgn}(a-b)}
{\mathbb I}
\otimes e_{aa}  \otimes {\mathbb I}\otimes {\mathbb I}\otimes  e_{bb}\otimes {\mathbb I}  \Bigg)\otimes {\mathbb I}\cdots, \nonumber\\
\rho_{a_3}({\mathbb U}_{l^{(3)}}) = \cdots {\mathbb I} \otimes \Bigg(\sum_{a \neq b}
{\mathbb I}  \otimes  {\mathbb I}\otimes e_{ab}\otimes {\mathbb I}  \otimes {\mathbb I}\otimes e_{ba}  \nonumber\\
\phantom{\rho_{a_3}({\mathbb U}_{l^{(3)}}) =}- \sum_{a\neq b}
a_3^{-{\rm sgn}(a-b)}
{\mathbb I}
\otimes  {\mathbb I} \otimes e_{aa} \otimes {\mathbb I}\otimes {\mathbb I} \otimes e_{bb}  \Bigg)\otimes {\mathbb I}\cdots  .
\end{gather*}
Our purpose is to show that
\[
\big[\Theta({\mathbb U}_l),  {\mathrm f}_{i}^{\otimes N}(\Delta(x))\big] =0, \qquad x \in U_{q_i}(sl_2),
\]
for $i=0,\dots,3$. Let us focus for simplicity, and for saving writing on $N =2$, since it is then easy to generalize for any $N$
(see also~\cite{doikoumartin}).
We shall show in detail the computation for the $f_0$ representation, the rest follow in the same spirit.
Consider f\/irst the element $e$ of the $U_{q_0}(sl_2)$ quantum algebra. The left-hand side of
$[\Theta({\mathbb U}_1),  f_0^{\otimes 2}(\Delta(e))]$ is written explicitly as
\begin{gather*}
\Theta({\mathbb U}_1) {\mathrm f}_0^{\otimes 2}(\Delta(e))
 = \Bigg  (\!\sum_{a \neq b}\!
e_{ab} \otimes {\mathbb I} \otimes {\mathbb I}\otimes e_{ba} \otimes {\mathbb I}\otimes {\mathbb I}  -
\sum_{a\neq b}\! a_1^{-{\rm sgn}(a-b)} e_{aa}
\otimes {\mathbb I} \otimes {\mathbb I}\otimes  e_{bb} \otimes  {\mathbb I}\otimes {\mathbb I}  \Bigg) \nonumber\\
\qquad {} \times\Bigg (\sum_{m \neq n}
{\mathbb I}  \otimes e_{mn} \otimes {\mathbb I}\otimes {\mathbb I}  \otimes e_{nm}\otimes {\mathbb I}  - \sum_{m\neq n}
a_2^{-{\rm sgn}(m-n)}{\mathbb I}
\otimes e_{mm}  \otimes {\mathbb I} \otimes {\mathbb I}\otimes  e_{nn}\otimes {\mathbb I} \Bigg ) \nonumber\\
\qquad {} \times\Bigg (\sum_{k \neq l}
{\mathbb I}  \otimes  {\mathbb I}\otimes e_{kl}\otimes {\mathbb I}  \otimes {\mathbb I}\otimes e_{lk}  -
\sum_{k\neq l}a_3^{-{\rm sgn}(k-l)} {\mathbb I}
\otimes  {\mathbb I} \otimes e_{kk} \otimes {\mathbb I}\otimes {\mathbb I} \otimes e_{ll} \Bigg) \nonumber\\
\qquad{} \times\Big (\big(q_0^{-{e_{11}\over 2}} \otimes e_{11} \otimes e_{11}+ q_0^{{e_{22} \over 2}} \otimes e_{22} \otimes e_{22} +
{\mathbb I} \otimes e_{11} \otimes _{22} + {\mathbb I} \otimes e_{22} \otimes e_{11}\big)\nonumber\\
\qquad {} \otimes e_{12} \otimes e_{12} \otimes e_{12}
+\otimes e_{12} \otimes e_{12} \otimes e_{12} \nonumber\\
\qquad   {}
\otimes\big(q_0^{{e_{11}\over 2}} \otimes e_{11} \otimes e_{11}+
q_0^{-{e_{22} \over 2}} \otimes e_{22} \otimes e_{22} +
{\mathbb I} \otimes e_{11} \otimes _{22} + {\mathbb I} \otimes e_{22} \otimes e_{11}\big)  \Big ).
\end{gather*}
Using the property $e_{ij}  e_{kl} = \delta_{jk}  e_{il}$ and after some quite cumbersome algebra we end up with
the following expression for the f\/irst
\begin{gather}
\Theta({\mathbb U}_1) {\mathrm f}_0^{\otimes 2}(\Delta(e)) =
q_0^{{1\over 2}} e_{12} \otimes e_{12}\otimes e_{12}
\otimes e_{22} \otimes e_{22} \otimes e_{22} \nonumber\\
\qquad{}+ q_0^{-{1\over 2}} e_{22} \otimes e_{22} \otimes e_{22} \otimes e_{12}
\otimes e_{12} \otimes e_{12}  -a_3^{-1}q_0^{{1\over 2}} e_{12} \otimes e_{12} \otimes e_{22} \otimes e_{22} \otimes e_{22} \otimes e_{12}
\nonumber\\
\qquad{} -
a_3 q_0^{-{1 \over2}}e_{22} \otimes e_{2} \otimes e_{12} \otimes e_{12} \otimes e_{12} \otimes e_{22}
 - a_2^{-1} q_0^{1\over 2} e_{12} \otimes e_{22} \otimes e_{12} \otimes e_{22} \otimes e_{12} \otimes e_{22}\nonumber\\
\qquad{} -
a_2 q_0^{-{1\over 2}}e_{22} \otimes e_{12} \otimes e_{22} \otimes e_{12} \otimes e_{22} \otimes e_{12}
+a_2^{-1} a_3^{-1} q_0^{1\over 2} e_{12} \otimes e_{22} \otimes e_{22} \otimes e_{22} \otimes e_{12} \otimes e_{12}
\nonumber\\
\qquad {}+
a_2 a_3 q_0^{-{1\over 2}} e_{22} \otimes e_{12} \otimes e_{12} \otimes e_{12} \otimes e_{22} \otimes e_{22} -
a_1^{-1} q_0^{1\over 2} e_{22} \otimes e_{12} \otimes e_{12} \otimes e_{12} \otimes e_{22} \otimes e_{22}\nonumber\\
\qquad{}  -
a_1q_0^{-{1\over 2}} e_{12} \otimes e_{22} \otimes e_{22} \otimes e_{22} \otimes e_{12} \otimes e_{12} +
a_1^{-1} a_3^{-1} q_0^{1\over 2} e_{22} \otimes e_{12} \otimes e_{22} \otimes e_{12} \otimes e_{22} \otimes e_{12}\nonumber\\
\qquad{} +
a_1 a_2 q_0^{-{1\over 2}} e_{12} \otimes e_{22} \otimes e_{12} \otimes e_{22} \otimes e_{12} \otimes e_{12}\nonumber\\
\qquad{} +
a_1^{-1} a_2^{-1} q_0^{{1\over 2}} e_{22} \otimes e_{22} \otimes e_{12} \otimes e_{12} \otimes e_{12} \otimes e_{22}\nonumber\\
\qquad{} +
a_1 a_2 q_0^{-{1\over 2}} e_{12} \otimes e_{12} \otimes e_{22} \otimes e_{22} \otimes e_{22} \otimes e_{12} \nonumber\\
\qquad{} -
a_1^{-1} a_2^{-1} a_3^{-1}q_0^{1\over 2} e_{22} \otimes e_{22} \otimes e_{22} \otimes e_{12} \otimes e_{12} \otimes e_{12}\nonumber\\
\qquad{}-
a_1 a_2 a_3 q_0^{-{1\over 2}} e_{12} \otimes e_{12} \otimes e_{12} \otimes e_{22} \otimes e_{22} \otimes e_{22}.
\label{LHSbig}
\end{gather}
and the second term of the commutator
\begin{gather}
{\mathrm f}_0^{\otimes 2}(\Delta(e)) \Theta({\mathbb U}_1) =
q_0^{-{1\over 2}} e_{12} \otimes e_{12} \otimes e_{12} \otimes e_{11} \otimes e_{11} \otimes e_{11} \nonumber\\
\qquad{} +
q_0^{1\over 2} e_{11} \otimes e_{11} \otimes e_{11} \otimes e_{12} \otimes e_{12} \otimes e_{12} -
 a_3 q_0^{-{1\over 2}} e_{12} \otimes e_{12} \otimes e_{11} \otimes e_{11} \otimes e_{11} \otimes e_{12} \nonumber\\
 \qquad{} -
a_3^{-1} q_0^{1\over 2}e_{11} \otimes e_{11} \otimes e_{12} \otimes e_{12} \otimes e_{12} \otimes e_{11}-
a_2 q_0^{-{1\over 2}} e_{12} \otimes e_{11} \otimes e_{12} \otimes e_{11} \otimes e_{12} \otimes e_{11}\nonumber\\
\qquad{} -
a_2^{-1} q_0^{1\over 2} e_{11} \otimes e_{12} \otimes e_{11} \otimes e_{12} \otimes e_{11} \otimes e_{12} +
 a_2 a_3 q_0^{-{1\over 2}} e_{12} \otimes e_{11} \otimes e_{11} \otimes e_{11} \otimes e_{12} \otimes e_{12} \nonumber\\
 \qquad{} +
a_2^{-1} a_3^{-1} q_0^{1\over 2} e_{ 11} \otimes e_{12} \otimes e_{12} \otimes e_{12} \otimes e_{11} \otimes e_{11} -
 a_1 q_0^{-{1\over 2}} e_{11} \otimes e_{12} \otimes e_{12} \otimes e_{12} \otimes e_{11} \otimes e_{11} \nonumber\\
 \qquad{} -
a_1^{-1} q_0^{1\over 2} e_{12} \otimes e_{11} \otimes e_{11} \otimes e_{11} \otimes e_{12} \otimes e_{12} +
 a_1 a_3 q_0^{-{1\over 2}} e_{11} \otimes e_{12} \otimes e_{11} \otimes e_{12} \otimes e_{11} \otimes e_{12} \nonumber\\
 \qquad{} +
a_1^{-1} a_3^{1} q_0^{1\over 2}  e_{12}  \otimes e_{11} \otimes e_{12} \otimes e_{11} \otimes e_{12} \otimes e_{11}\! +\!
 a_1 a_2 q_0^{-{1\over 2}}  e_{11} \otimes e_{11} \otimes e_{12} \otimes e_{12} \otimes e_{12} \otimes e_{11} \nonumber\\
 \qquad{}  +
a_1^{-1}a_2^{-1} q_0^{1\over 2} e_{12} \otimes e_{12} \otimes e_{11} \otimes e_{11} \otimes e_{11} \otimes e_{12}  \nonumber\\
\qquad{} - a_1 a_2 a_2 q_0^{-{1\over 2}} e_{11} \otimes e_{11} \otimes e_{11} \otimes e_{12} \otimes e_{12} \otimes e_{12}\nonumber\\
\qquad{}  -
a_1^{-1} a_2^{-1} a_3^{-1} q_0^{1\over 2} e_{12} \otimes e_{12} \otimes e_{12} \otimes e_{11} \otimes e_{11} \otimes e_{11}.
\label{RHSbig}
\end{gather}
Now
notice that every two terms in each side of the commutator are canceled provided that $q_0 =  a_1  a_2  a_3$. This result has also been found to hold for the $n=4, 5$ cases; each side of
the commutator vanishes identically, for the appropriate form of $q_i$, $q_{i_1 i_2}$.

The commutator of the junction representation with ${\mathrm f}_0^{\otimes 2}(\Delta(f))$ follows in the
same spirit and the expressions obtained are very similar to those presented in \eqref{LHSbig}
and \eqref{RHSbig}. However, the commutator of the representation with ${\mathrm f}_0^{\otimes 2}(\Delta(h))$ is even easier to cope with, since it gives identically zero, without
requiring any particular form for the~$q_i$. This has also been outlined in the general
comments of Section~\ref{section2.2}, where the relation $[\Theta({\mathbb U}_1),  {\mathrm f}_{i_1\dots i_m}^{\otimes N}(\Delta^{(N)}(h))]=0$ has been proved there by using some simple heuristic arguments. In the same spirit the proof can be extended for $i=1,  2,  3$ ($n=3$).

\section[Explicit expressions for $n=4$]{Explicit expressions for $\boldsymbol{n=4}$}\label{appendixC}

We present in this Appendix explicit expressions of the representations for the $n=4$ case
to further illustrate the structure of these representations.
We also write down the explicit value of $q_{i}$, $q_{ij}$ so that the respective representation is a quantum symmetry of our algebra
\begin{gather*}
{\mathrm f}_{0}(e)=e_{12}\otimes e_{12}\otimes e_{12}\otimes e_{12}, \qquad {\mathrm f}_{0}(f) =
 e_{21}\otimes e_{21}\otimes e_{21}\otimes e_{21},
\\
{\mathrm f}_{0}(h)=e_{11}\otimes e_{11}\otimes e_{11}\otimes e_{11}-
e_{22}\otimes e_{22}\otimes e_{22}\otimes e_{22},\qquad q_0=-a_1a_2a_3a_4,  \\
{\mathrm f}_{1}(e)=e_{21}\otimes e_{12}\otimes e_{12}\otimes e_{12}, \qquad {\mathrm f}_{1}(f) =
 e_{12}\otimes e_{21}\otimes e_{21}\otimes e_{21}, \nonumber\\
{\mathrm f}_{1}(h)=e_{22}\otimes e_{11}\otimes e_{11}\otimes e_{11}-
e_{11}\otimes e_{22}\otimes e_{22}\otimes e_{22},\qquad q_1=-a_1^{-1}a_2a_3a_4,\\
{\mathrm f}_{2}(e)=e_{12}\otimes e_{21}\otimes e_{12}\otimes e_{12}, \qquad {\mathrm f}_{2}(f) =
 e_{21}\otimes e_{12}\otimes e_{21}\otimes e_{21},
 \\
{\mathrm f}_{2}(h)=e_{11}\otimes e_{22}\otimes e_{11}\otimes e_{11}-
e_{22}\otimes e_{11}\otimes e_{22}\otimes e_{22},\qquad q_2=-a_1a_2^{-1}a_3a_4,\\
{\mathrm f}_{3}(e)=e_{12}\otimes e_{12}\otimes e_{21}\otimes e_{12}, \qquad {\mathrm f}_{3}(f) =
 e_{21}\otimes e_{21}\otimes e_{12}\otimes e_{21},
 \\
{\mathrm f}_{3}(h)=e_{11}\otimes e_{11}\otimes e_{22}\otimes e_{11}-
e_{22}\otimes e_{22}\otimes e_{11}\otimes e_{22},\qquad q_3=-a_1a_2a_3^{-1}a_4,\\
{\mathrm f}_{4}(e)=e_{12}\otimes e_{12}\otimes e_{12}\otimes e_{21}, \qquad {\mathrm f}_{4}(f) =
 e_{21}\otimes e_{21}\otimes e_{21}\otimes e_{12},
 \\
{\mathrm f}_{4}(h)=e_{11}\otimes e_{11}\otimes e_{11}\otimes e_{22}-
e_{22}\otimes e_{22}\otimes e_{22}\otimes e_{11},\qquad q_4=-a_1a_2a_3a_4^{-1},\\
{\mathrm f}_{34}(e)=e_{12}\otimes e_{12}\otimes e_{21}\otimes e_{21}, \qquad {\mathrm f}_{34}(f) =
 e_{21}\otimes e_{21}\otimes e_{12}\otimes e_{12},
 \\
{\mathrm f}_{34}(h)=e_{11}\otimes e_{11}\otimes e_{22}\otimes e_{22}-
e_{22}\otimes e_{22}\otimes e_{11}\otimes e_{11},\qquad q_{34}=-a_1a_2a_3^{-1}a_4^{-1},\\
{\mathrm f}_{24}(e)=e_{12}\otimes e_{21}\otimes e_{12}\otimes e_{21}, \qquad {\mathrm f}_{24}(f) =
 e_{21}\otimes e_{12}\otimes e_{21}\otimes e_{12},
 \\
{\mathrm f}_{24}(h)=e_{11}\otimes e_{22}\otimes e_{11}\otimes e_{22}-
e_{22}\otimes e_{11}\otimes e_{22}\otimes e_{11} ~~~~~q_{24}=-a_1a_2^{-1}a_3a_4^{-1}, \\
{\mathrm f}_{23}(e)=e_{12}\otimes e_{21}\otimes e_{21}\otimes e_{12}, \qquad {\mathrm f}_{23}(f) =
 e_{21}\otimes e_{12}\otimes e_{12}\otimes e_{21},
 \\
{\mathrm f}_{23}(h)=e_{11}\otimes e_{22}\otimes e_{22}\otimes e_{11}-
e_{22}\otimes e_{11}\otimes e_{11}\otimes e_{22}, \qquad q_{23}=-a_1a_2^{-1}a_3^{-1}a_4.
\end{gather*}

\subsection*{Acknowledgements}

NK acknowledges f\/inancial support provided by the Research Committee of the University of Patras
via a K.Karatheodori fellowship under contract number C.915, and partial support by the LLP/Erasmus
Placements 2009-2010 program with contract 0099/2009. He would also like to thank the CPhT of Ecole Polytechnique
for kind hospitality and partial support by the ERC Advanced Grant 226371, the ITN programme PITN- GA-2009-237920
and the IFCPAR CEFIPRA programme 4104-2 during the completion of this work.

\pdfbookmark[1]{References}{ref}
\LastPageEnding

\end{document}